\documentclass[twocolumn,showpacs,aps,prd,amsmath,amssymb]{revtex4-1}

\usepackage[dvips]{graphicx}
\usepackage{subfigure}
\usepackage{color}
\usepackage{epsfig}
\usepackage{indentfirst}
\usepackage{multirow}
\usepackage{amsmath}
\usepackage{amssymb}
\usepackage{bm}
\usepackage{fancyhdr}
\usepackage{enumerate}
\usepackage[perpage,symbol]{footmisc}
\usepackage[bookmarksnumbered,pdfstartview=FitH,colorlinks=true,menucolor=black,linkcolor=blue]{hyperref}
\usepackage{xspace}

\newcommand{\jpsi}{\mbox{$J/\psi$}~}

\newcommand{\dswd}{\mbox{$J/\psi \to D^{-}_{s}e^{+}\nu_{e}$}~}
\newcommand{\dsstarwd}{\mbox{$J/\psi \to {D^{*}_{s}}^{-}e^{+}\nu_{e}$}~}
\newcommand{\dsordsstarwd}{\mbox{$J/\psi \to D^{(*)-}_{s}e^{+}\nu_{e}$}~}
\newcommand{\Umiss}{\ensuremath{U_\mathrm{miss}}\xspace}

\begin{document}

\title{%
 \boldmath Search for the weak decays $J/\psi \to D^{(*)-}_{s} e^+ \nu_{e}+c.c.$
}

\author{
  \begin{small}
    \begin{center}
      M.~Ablikim$^{1}$, M.~N.~Achasov$^{8,a}$, X.~C.~Ai$^{1}$, O.~Albayrak$^{4}$, M.~Albrecht$^{3}$, 
      D.~J.~Ambrose$^{43}$, A.~Amoroso$^{47A,47C}$, F.~F.~An$^{1}$, Q.~An$^{44}$, J.~Z.~Bai$^{1}$
% ...
      \\
      \vspace{0.2cm}
      (BESIII Collaboration)\\
      \vspace{0.2cm} 
      { \it
          $^{1}$ Institute of High Energy Physics, Beijing 100049, People's Republic of China\\
% ...
$^{3}$ Bochum Ruhr-University, D-44780 Bochum, Germany\\
          $^{4}$ Carnegie Mellon University, Pittsburgh, Pennsylvania 15213, USA\\
% ...
$^{8}$ G.I. Budker Institute of Nuclear Physics SB RAS (BINP), Novosibirsk 630090, Russia\\
% ...
$^{43}$ University of Rochester, Rochester, New York 14627, USA\\
          $^{44}$ University of Science and Technology of China, Hefei 230026, People's Republic of China\\
% ...
$^{47}$ (A)University of Turin, I-10125, Turin, Italy; (B)University of Eastern Piedmont, I-15121, Alessandria, Italy; (C)INFN, I-10125, Turin, Italy\\
% ...
$^{a}$ Also at the Novosibirsk State University, Novosibirsk, 630090, Russia\\
% ...
      }\end{center}
    \vspace{0.4cm}
  \end{small}
}

\affiliation{}

\date{\today}

\begin{abstract}
Using a sample of $2.25\times 10^8$ \jpsi events collected with the
BESIII detector at the BEPCII collider, we search for the
\jpsi semileptonic weak decay $J/\psi \to D^{-}_{s}
e^{+}\nu_{e}+c.c.$ with a much higher sensitivity than
previous searches. We also perform the first search for $J/\psi \to
D^{*-}_{s} e^{+}\nu_{e}+c.c.$ 
No significant excess of a signal above
background is observed in either channel. At the $90\%$ confidence level, the upper
limits are determined to be $\mathcal{B}(J/\psi \to
D^{-}_{s}e^{+}\nu_{e}+c.c.)<1.3\times10^{-6}$ and
$\mathcal{B}(J/\psi \to
{D^{*}_{s}}^{-}e^{+}\nu_{e}+c.c.)<1.8\times10^{-6}$, respectively.
Both are consistent with Standard Model predictions.

\end{abstract}


\maketitle
\pagenumbering{arabic}

\section{Introduction}
\label{sec:introduction}

\begin{figure}[htbp]
\includegraphics[width=0.35\textwidth]{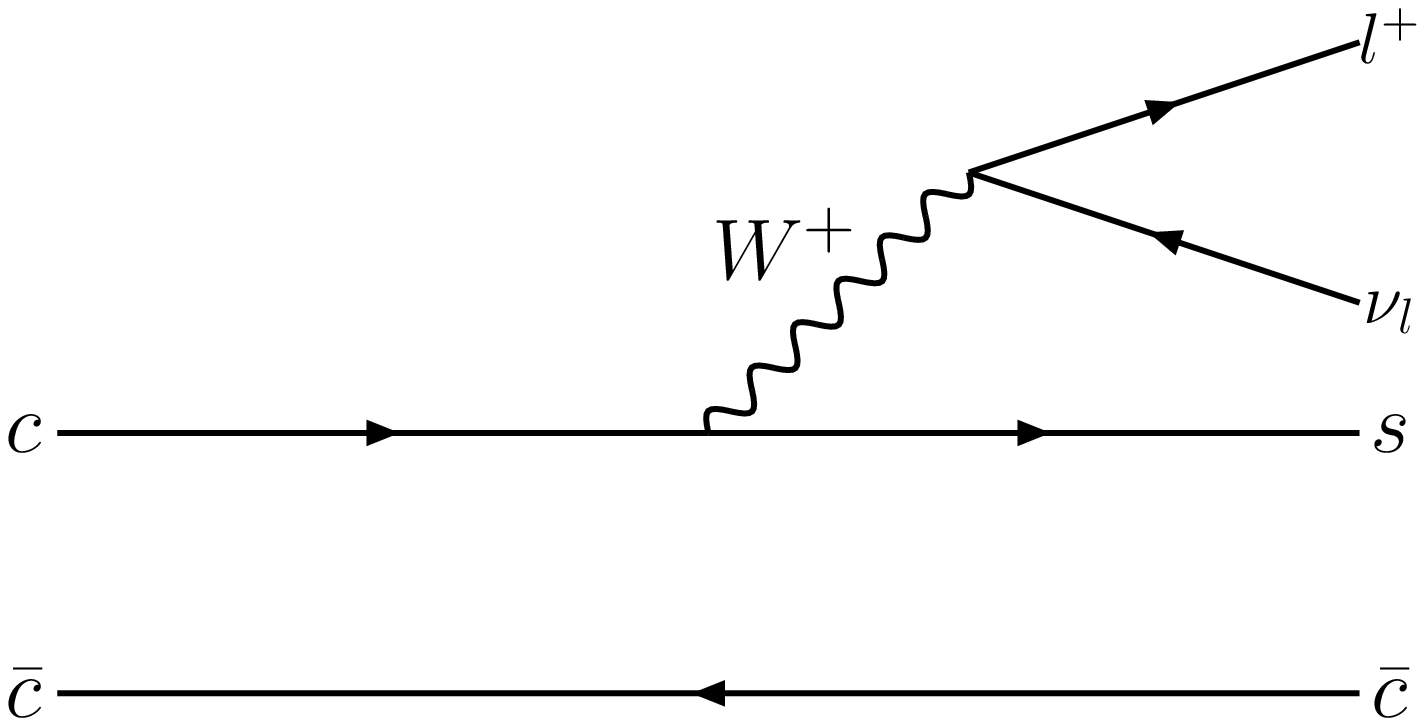}
\caption{Feynman diagrams for $J/\psi \to D^{(*)-}_{s} l^{+}\nu_{l}$ at the tree level.}
\label{fig:feynman}
\end{figure}

The $J/\psi$ particle, lying below the open charm threshold, cannot
decay into a pair of charmed mesons. However, the $J/\psi$ can decay
into a single charmed meson via the weak
interaction~\cite{Sanchis:1992pv}. Weak decays of the $J/\psi$ are
rare processes, and the inclusive branching fractions of $J/\psi$
decays to single $D$ or $D_s$ mesons are predicted to be of the order
of $10^{-8}$ or less~\cite{SanchisLozano:1993ki} in the Standard
Model (SM). Figure.~\ref{fig:feynman} shows the tree-level Feynman
diagram within the SM for the decays $J/\psi \to D_{s}^{(*)}l\nu$
($l=e$ or $\mu$). Most recent theoretical calculations predict the
$J/\psi \to D_{s}^{(*)}l\nu$ branching fractions to be $\simeq
10^{-10}$ by using QCD sum rules and employing the covariant
light-front quark model~\cite{Wang:2007ys}. However, as mentioned in
Refs.~\cite{xm.zhang2001,Li:2012vk,Datta:1998yq,hill:1995}, the
branching fractions of $J/\psi \to D(\bar{D}) X$ (with $X$ denoting any
hadrons) could be enhanced when new interaction couplings are
considered, such as in the top-color models, the minimal
super-symmetric SM with R-parity violation, or the
two-Higgs-doublet model. It is interesting to note that the ratio
between $J/\psi \to D_{s}^{*} l\nu$ and $D_{s} l\nu$ is predicted to
be $1.5 \sim 3.1$ in Ref.~\cite{SanchisLozano:1993ki,Wang:2007ys},
where part of the theoretical uncertainties cancel.

The BES collaboration has studied several weak decays,
including semileptonic and nonleptonic weak decays of the $J/\psi$.
With a $5.8\times10^7$ $J/\psi$ events sample, the upper limit for
$\mathcal{B}(J/\psi \to D^{-}_{s}e^{+}\nu_{e}+c.c.)$ was found to be $3.6\times
10^{-5}$ at the $90\%$ C.L.~\cite{Ablikim:2006qt},
while the $J/\psi \to D_s^{*-} e^+ \nu_e+c.c.$ has never been
studied in experiments before. When we refer to $+c.c.$, we mean 
the combination of $J/\psi \to D_s^{(*)-} e^+ \nu_e$ and the charge conjugated
modes $J/\psi \to D_s^{(*)+} e^- \bar{\nu_e}$. 
In the following, the signals are the sum of both
modes and charge conjugation is implied unless otherwise specified.
Using a sample of $2.25\times 10^8$ \jpsi events
collected with the BESIII detector at the Beijing Electron Positron
Collider (BEPCII)~\cite{Ablikim:2012cn}, we search for the weak
decays $J/\psi \to D_s^- e^+ \nu_e$ and $J/\psi \to D_s^{*-} e^+
\nu_e$. The $D_s^-$ meson is reconstructed via four decay modes:
$D_{s}^{-} \to K^{+} K^{-} \pi^{-}$, $D_{s}^{-} \to K^{+} K^{-}
\pi^{-} \pi^{0}$, $D_{s}^{-} \to K_S^0 K^{-}$,
and $D_{s}^{-} \to K_S^0 K^{+}\pi^{-} \pi^{-}$, where the
$\pi^0$ and $K_S^0$ mesons are reconstructed from
their $\gamma \gamma$ and $\pi^+\pi^-$ decays, respectively. The $D_{s}^{*}$
candidate is reconstructed from its radiative transitions to $D_s$.
A $478\ \mathrm{pb}^{-1}$ data sample collected at the center-of-mass energy $\sqrt{s}=4.009\
\mathrm{GeV}$~\cite{Ablikim:2012ht} is used to study
systematic uncertainties.

\section{BESIII Experiment}
\label{sec:detector} The BESIII detector is a magnetic
spectrometer~\cite{Ablikim:2009aa} located at BEPCII, which is a
double-ring $e^+ e^-$ collider with a design peak luminosity of
$10^{33}$ cm$^{-2}$ s$^{-1}$ at a center-of-mass energy of 3.773
GeV. The cylindrical core of the BESIII detector consists of a
helium-based main drift chamber (MDC), a plastic scintillator
time-of-flight system (TOF), and a CsI (Tl) electromagnetic
calorimeter (EMC), which are all enclosed in a superconducting
solenoidal magnet providing a 1.0 T magnetic field.  The solenoid is
supported by an octagonal flux-return yoke with modules of resistive
plate muon counters interleaved with steel. The acceptance for
charged particles and photons is 93\% over a 4$\pi$ solid angle. The
momentum resolution for a charged particle at 1 GeV/c is 0.5\%, and
the ionization energy loss per unit path-length ($dE/dx$) resolution is 6\%. The EMC measures photon energies with
a resolution of 2.5\% (5\%) at 1 GeV in the barrel (end caps). The
time resolution for the TOF is 80 ps in the barrel and 110 ps in the
end caps.

Monte Carlo (MC) simulations are used to determine the detection efficiency, study
backgrounds and optimize event selection criteria. A {\sc
  geant4}-based \cite{Agostinelli:2002hh} simulation is
used to simulate the BESIII detector response. Electron-positron annihilation into a $J/\psi$
resonance is simulated at energies around $\sqrt{s}=3.097\
\mathrm{GeV}$, while the beam energy and its energy spread are set according to
measurements of the Beam Energy Measurement System~\cite{Abakumova:2011rp}. The
production of the $J/\psi$ resonance is implemented with the generator {\sc
kkmc}~\cite{kkmc}.
The signal channels are generated with
a new generator implemented in {\sc evtgen}~\cite{evtgen}, and we assume the process
$J/\psi \to D^{(*)-}_s e^+\nu_e$ is dominated by the weak interaction, i.e. via
the $c \to s$ charged current process, while the effects
of hadronization and quark spin flip are ignored.
The known decay modes of the $J/\psi$ resonance are generated by {\sc
evtgen}~\cite{evtgen} with branching fractions set according to the world average values of the
Particle Data Group~\cite{Beringer:1900zz}, while the unknown decays are generated
by {\sc lundcharm}~\cite{lundcharm}.  
A sample of $2.25\times 10^8$ generic
$J/\psi$ decays ("inclusive MC") is used to identify potential background
channels.

\section{Event Selection and Data Analysis}
\label{sec:selection}

Tracks from charged particles are reconstructed using hit information from the MDC.
We select tracks in the polar angle range $|\cos\theta| < 0.93$ and
require that they pass within $\pm 10$ cm from the interaction point
(IP) along the beam and within $\pm 1$ cm transverse to the beam
direction. The charged particle identification (PID) is based on a
combination of $dE/dx$ and TOF information, and the probability of
each particle hypothesis ($P(i)$ with $i=e/\pi/K$)
is calculated. A pion candidate is required to satisfy
$P(\pi)>0.001$ and $P(\pi)>P(K)$; for kaons, $P(K)>0.001$ and
$P(K)>P(\pi)$ are required; and for electrons or positrons, we
require the track from charged particles to satisfy $P(e)>0.001$ and $P(e)>P(K)$ and
$P(e)>P(\pi)$ as well as $0.80<E/p<1.05$, where $E/p$ is the ratio
of the energy deposited in the EMC to the momentum of the track
measured by the MDC.

The $K_S^0$ candidates are reconstructed from pairs of
oppositely charged tracks, which are assumed to be pions without a PID
requirement, and where the IP requirements are relaxed to $20$~$\mathrm{cm}$ in the direction along the beam. For each pair of tracks, a
primary vertex fit and a secondary vertex fit are performed and
the $K_S^0$ decay length is required to be two times larger than its
fit error.
The resulting track parameters from the secondary vertex fit are
used to calculate the invariant mass $M(\pi^+\pi^-)$.
The $\pi^+\pi^-$ combinations with an invariant
mass $0.487 \  \mathrm{GeV/c^{2}} <M(\pi^+\pi^-)<0.511 \
\mathrm{GeV/c^{2}}$ are kept as $K_S^0$ candidates.
Multiple $K_S^0$ candidates are allowed in one event.

Photon candidates are reconstructed based on the showers in both the
EMC barrel region ($|\cos\theta|<0.8$) and the end cap regions
($0.86<|\cos\theta|<0.92$). Showers from the barrel region must have
a minimum energy of $25 \ \mathrm{MeV}$, while those in the end caps
must have at least  $50 \ \mathrm{MeV}$.  To exclude showers from
charged particles, a photon candidate must be separated by at least
20$^\circ$ from any charged particle track with respect to the
interaction point. The EMC timing information ($0
\ \mathrm{ns} \leqslant T  \leqslant 700\ \mathrm{ns}$) is used to further
suppress electronic noise and energy depositions unrelated to the
event.

The $\pi^0$ candidates are reconstructed from pairs of photons.  A kinematic fit is
performed constraining the invariant mass of the photon pair to the known $\pi^0$
mass~\cite{Beringer:1900zz}. The combination with the minimum $\chi^2$ from the kinematic fit that satisfies $\chi^2 <
100$, and $0.115 \ \mathrm{GeV/c^{2}} <M(\gamma\gamma)<0.150\ \mathrm{GeV/c^{2}}$ is kept for
further analysis. The $\pi^0$ candidates with both photons from the end cap regions are
excluded due to poor resolution in this region of the detector.

With the previously described charged and neutral particle
candidates, the $D^-_{s}$ candidates can be reconstructed through
the four decay modes mentioned in the Introduction; we name them
$KK\pi$, $KK\pi\pi$, $K_S^0K$ and $K_S^0K\pi\pi$, and number each as
the $k$th ($k=1...4$) decay mode, in sequence. Since the 
resolution of the reconstructed $D^-_{s}$ mass is different for each decay mode,
the invariant mass of $D^-_{s}$ candidates is required to be in
different mass windows, which are taken as three
times the respective resolution ($\pm 3 \sigma$ around its central value). For 
\dsstarwd, the $D_s^-$ and an additional photon candidate are
combined to reconstruct $D^{*-}_s$ candidates, and  the invariant
mass difference ($\Delta M$) between $D^-_{s} \gamma$ and $D^-_{s}$
is required to satisfy $0.125 \ \mathrm{GeV/c^{2}} <\Delta M<0.150\
\ \mathrm{GeV/c^{2}}$. To avoid bias, we set no requirement
to select the best $D^-_{s}$ or $D^{*-}_s$ candidate, and multiple
$D^-_{s}$ or $D_s^{*-}$ candidates are allowed in one event.
According to the MC simulations, after all selection criteria are
applied, events with multiple candidates occur in about $0.1\%$ cases for each
mode in $J/\psi \to D_s^- e^+ \nu_e$ and about $0.2\%$ for each mode
in $J/\psi \to D_s^{*-} e^+ \nu_e$. For real data, only a few events
are observed and no events with multiple candidates are found, so the effect of
the multiplicity of candidates can be safely ignored.

Once a $D^-_{s}$ or $D^{*-}_{s}$ is reconstructed, the signal event
candidate is required to contain a positron track. Events that
include charged particles other than those from the $D_s^-$ and the
positron candidate are vetoed. To reduce background contributions from
misidentified events with extra photons, we require the total
energy of those extra neutral particles be less than $0.2$ or
$0.3$ GeV for $D_s^-$ or $D^{*-}_s$ in the modes of $K^+K^-\pi^-$,
$K_S^0K^-$ and $K_S^0K^+\pi^-\pi^-$, respectively, and $0.15$ or
$0.2$ GeV for the $K^+K^-\pi^-\pi^0$ mode.
These selection criteria are chosen
by optimizing the ratio $S/\sqrt{B}$, where $S$ and $B$ are the numbers of
signal events from the signal MC sample and expected background events from
the inclusive MC sample, respectively.

\begin{figure*}[htbp]
\includegraphics[width=0.8\textwidth]{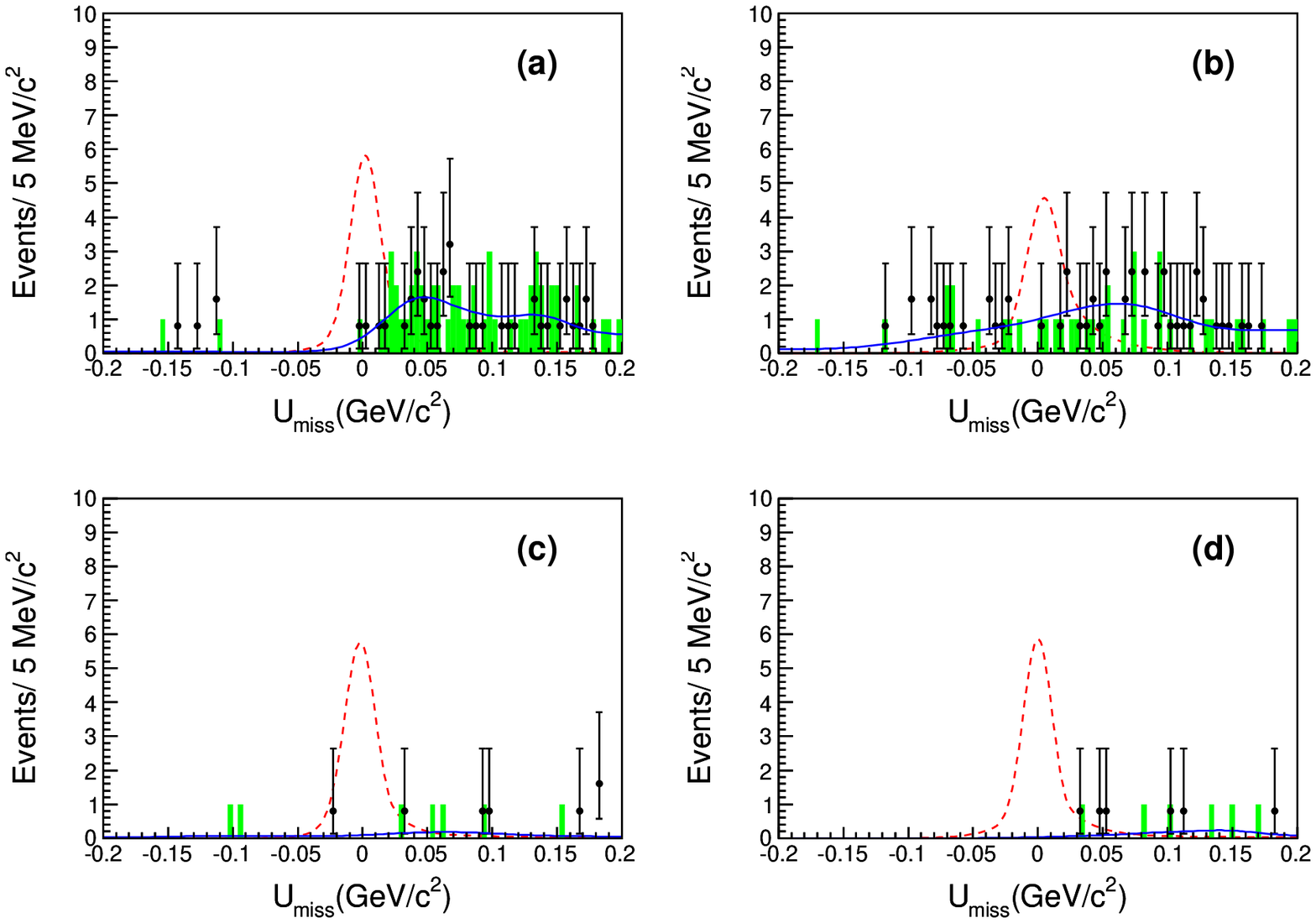}
\caption{$\Umiss$ distributions for  \dswd: (a) $D_{s}^{-} \to
K^{+} K^{-}
  \pi^{-}$; (b) $D_{s}^{-} \to K^{+} K^{-} \pi^{-} \pi^{0}$; (c)  $D_{s}^{-}
  \to K^0_S K^{-}$; (d) $D_{s}^{-} \to K^0_S K^{+} \pi^{-} \pi^{-}$. 
   Data are shown by dots with error bars, the signal shapes 
   are shown with dashed curves, the background contributions 
   from inclusive MC simulations are shown with filled histograms, 
   and the results of simultaneous fit are shown with solid curves. 
   Here the signal shape is drawn with arbitrary normalization, 
   while the shapes of inclusive MC and fit are normalized to the data luminosity.}
\label{fig:dsumissdata}
\end{figure*}

\begin{figure*}[htbp]
\includegraphics[width=0.8\textwidth]{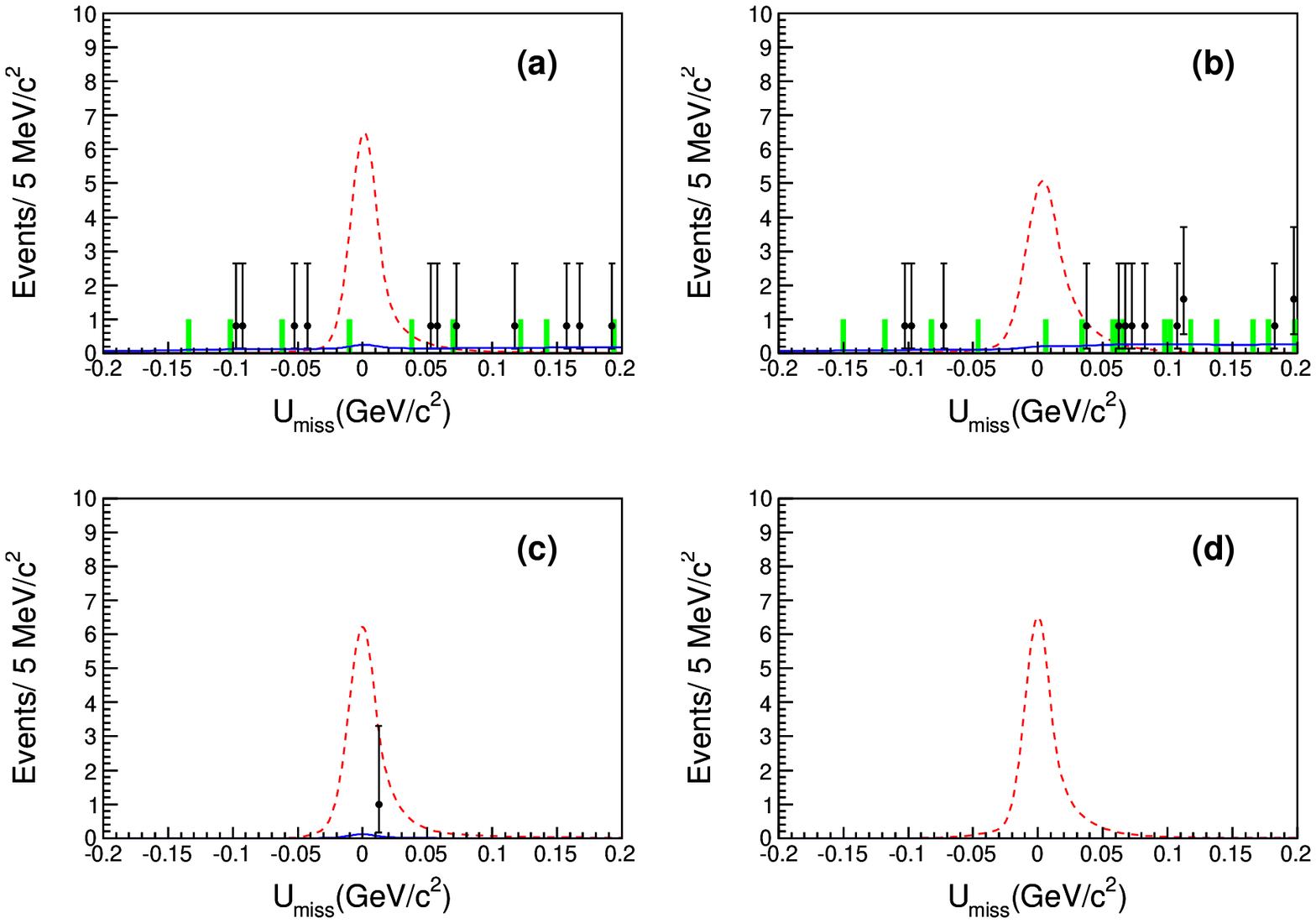}
\caption{$\Umiss$ distributions for \dsstarwd: (a) $D_{s}^{-} \to
K^{+} K^{-} \pi^{-}$; (b)  $D_{s}^{-} \to K^{+} K^{-} \pi^{-} \pi^{0}$; (c)
  $D_{s}^{-} \to K^0_S K^{-}$; (d)  $D_{s}^{-} \to K^0_S K^{+} \pi^{-}
  \pi^{-}$. 
   Data are shown by dots with error bars, the signal shapes 
   are shown with dashed curves, the background contributions 
   from inclusive MC simulations are shown with filled histograms, 
   and the results of simultaneous fit are shown with solid curves. 
   Here the signal shape is drawn with arbitrary normalization, 
   while the shapes of inclusive MC and fit are normalized with to the
   data luminosity.}
\label{fig:dsrumissdata}
\end{figure*}

For a $J/\psi \to D_s^{(*)-} e^+ \nu_e$ candidate, the undetected
neutrino leads to a missing energy $E_{miss} =
E_{J/\psi}-E_{D_s^{(*)-}}-E_{e^+}$ and a missing momentum
$\vec{p}_{miss} =
\vec{p}_{J/\psi}-\vec{p}_{D_{s}^{(*)-}}-\vec{p}_{e^+}$, where
$E_{D_s^{(*)-}}$ and $\vec{p}_{D_{s}^{(*)-}}$ ($E_{e^+}$ and
$\vec{p}_{e^+}$) are the energy and momentum of the $D_s^{(*)-}$
(positron).
We require
$|\vec{p}_{miss}|$ to be larger than $50\ \mathrm{MeV}$ to
suppress the background contributions from \jpsi hadronic decays in which a
pion is misidentified as a positron. The $J/\psi$
semileptonic decay events are extracted using the variable $\Umiss =
E_{miss}-|\vec{p}_{miss}|$. If the decay products of the $J/\psi$
semileptonic decay have been correctly identified, $\Umiss$ is
expected to peak around zero.  The
$\Umiss$ distributions of \dswd and \dsstarwd candidates are shown
in Figs.~\ref{fig:dsumissdata} and~\ref{fig:dsrumissdata},
respectively. The signal shapes obtained from MC simulations are
shown with dashed curves. No significant excess of signal
above background is observed in either mode.

From a MC study, we find that background events are mostly from
those decay modes where a pion is misidentified as an
electron/positron. For example, the process $J/\psi \to K^+ K^-
\pi^- \pi^+$ would be one potential background of $J/\psi \to
D_s^{-} e^+ \nu_e$, $D_s^- \to K^+ K^- \pi^-$. Background channels
from inclusive MC simulations are shown in
Figs.~\ref{fig:dsumissdata} and~\ref{fig:dsrumissdata} with filled
histograms. No peaking background is found, and the expected
background from MC is consistent with data. 

For each $D^-_s$ decay mode, $100,000$ exclusive signal MC events are generated,
and the detection efficiencies are determined to be $(24.46 \pm 0.17)\%$, $(11.08 \pm 0.13)\%$,
$(29.90 \pm 0.19)\%$ and $(13.74 \pm 0.12)\%$ for $KK\pi$, $KK\pi\pi^0$, $K_S^0K$ and 
$K_S^0K\pi\pi$ modes of \dswd, and $(16.59 \pm 0.17)\%$, $(7.40 \pm 0.15)\%$, $(19.62 \pm 0.17)\%$
and $(8.20 \pm 0.11)\%$ for $KK\pi$, $KK\pi\pi^0$, $K_S^0K$ and
$K_S^0K\pi\pi$ modes of \dsstarwd, respectively.

A simultaneous unbinned maximum likelihood fit is used to determine the event yields
of the four $D_s$ decay modes. The Bayesian
method~\cite{Beringer:1900zz} with a uniform prior is used to estimate the upper limits on the
number of signal events since no significant signals are observed for either $J/\psi$ weak decay
mode. We choose $-0.2\mathrm{\ GeV/c^2}  < \Umiss <0.2 \mathrm{\ GeV/c^2}$ as the fitting range.
The signal events are described by a sum of a Gaussian and a Crystal Ball
function~\cite{Gaiser:1982yw} with the parameters obtained from a fit to
the signal MC sample. The background shape is obtained from the inclusive
$J/\psi$ MC sample and modeled with a probability density function that
represents the shape of an external unbinned data set as a superposition of Gaussians~\cite{Cranmer:2000du}.
The likelihood for the $k$th $D^-_s$ decay
mode is constructed as
\begin{eqnarray}
  \mathcal{L}_k =  \prod^{N_k}_{i=1}\frac{N_\mathrm{total}{\cal B}_k\epsilon_k \mathcal{P}^\mathrm{sig}_{i,k}
  + N^\mathrm{bkg}_k\mathcal{P}^\mathrm{bkg}_{i,k}}{N_\mathrm{total}{\cal B}_k\epsilon_k + N^\mathrm{bkg}_k}\ ,
 \label{eq:likelihood}
\end{eqnarray}
where $N_\mathrm{total}$ is the total number of produced $J/\psi \to D^{(*)-}_{s}e^{+}\nu_{e}$
events in data,
$\mathcal{B}_k$ is the world average branching fraction of the
$k$th $D^-_s$ decay mode~\cite{Beringer:1900zz}, $\epsilon_k$ is
the detection efficiency of the $k$th $D^-_s$ decay mode, and
$N^\mathrm{bkg}_k$ is the number of background events in the $k$th $D^-_s$
decay mode. $N_k$ is the total number of selected events in the fit
region for the $k$th $D_s^-$ decay mode. 
$\mathcal{P}^\mathrm{sig}_{i,k}$ is the probability density function
of signal for the $k$th $D_s^-$ decay mode evaluated at the $i$th event;
similarly, $\mathcal{P}^\mathrm{bkg}_{i,k}$ is that of background.
The total likelihood $\mathcal{L}$ is the product of likelihoods for each
$D^-_s$ decay mode. A simultaneous unbinned
fit with floating amplitudes of signal and background is performed. 
No significant signal is found
by the fit as expected, and the fitting results are shown in the
Figs.~\ref{fig:dsumissdata} and~\ref{fig:dsrumissdata} as solid
curves.

We calculate the 90\% C.L. upper limit yield from
the fit, $N_\mathrm{total}^\mathrm{up}$, using
\begin{equation}
\frac{\int^{N_\mathrm{total}^\mathrm{up}}_0 \mathcal{L} (N_\mathrm{total})
dN_\mathrm{total}}{\int^{\infty}_0 \mathcal{L} (N_\mathrm{total})dN_\mathrm{total}}=0.90 \ ,
 \label{eq:90likelihood}
\end{equation}
where $\mathcal{L}(N_\mathrm{total})$ is the total likelihood  $\mathcal{L}$
at fixed $N_\mathrm{total}$.

\begin{figure}[htbp]
\subfigure{ \includegraphics[width=0.4\textwidth]{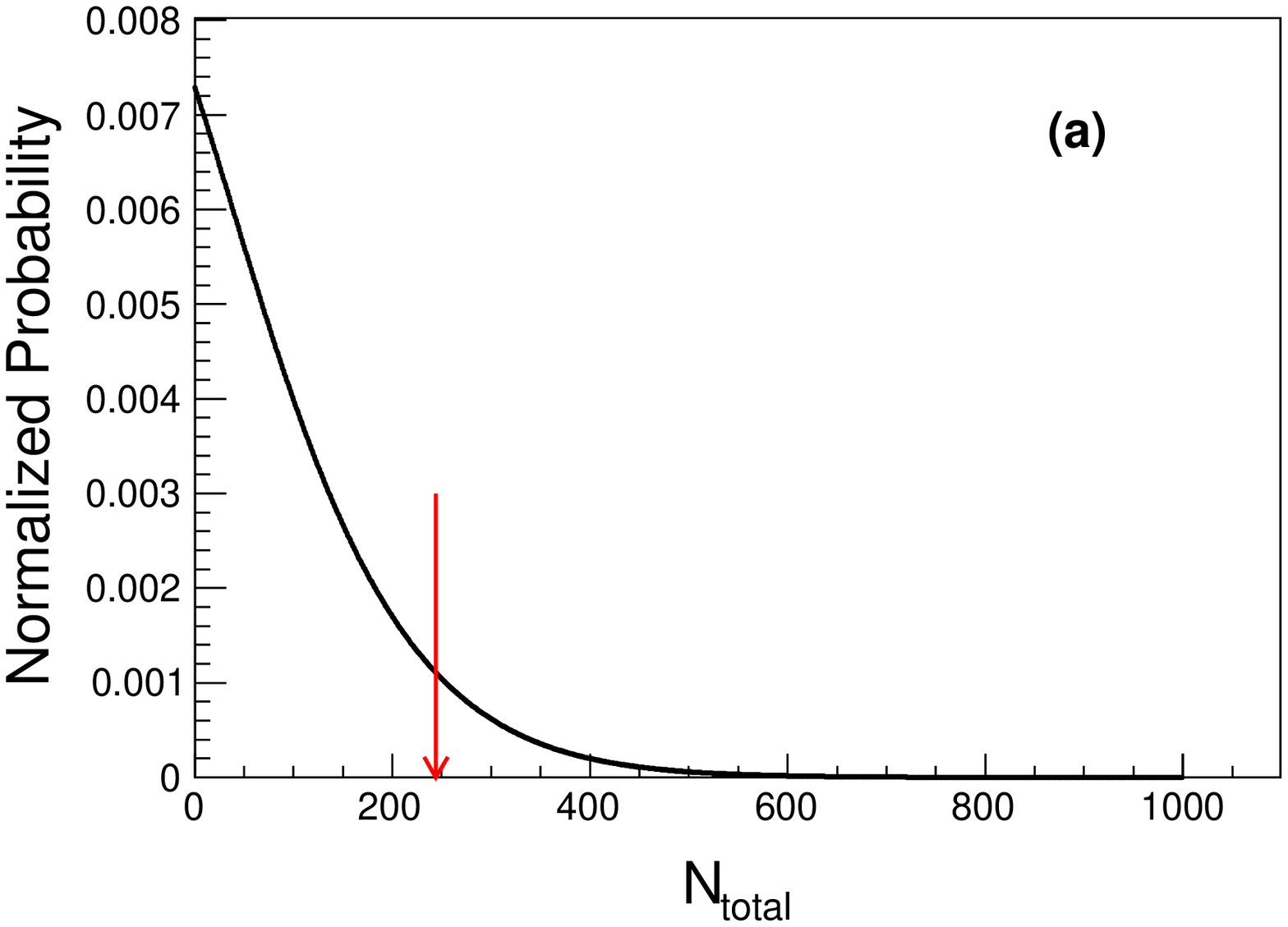} } \subfigure{
\includegraphics[width=0.4\textwidth]{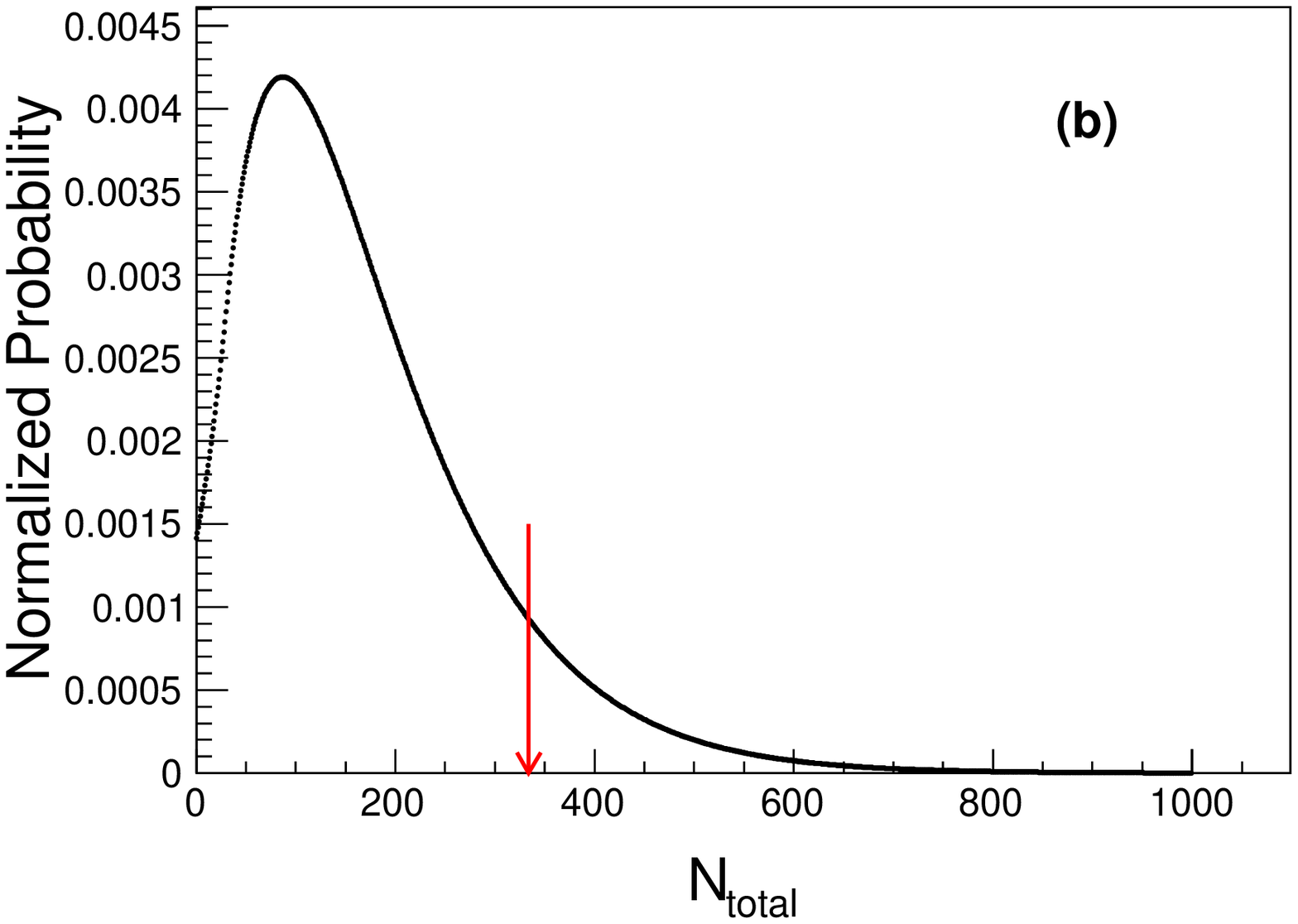} }
\caption{Normalized probabilities as a function of $N_\mathrm{total}$ in the (a) \dswd
and (b) \dsstarwd decay modes.  The red arrows indicate where 90\%
of the area is accumulated below the curves.} \label{fig:lh}
\end{figure}

In each fit, the likelihood value is obtained and the corresponding
probabilities are calculated as shown in Fig.~\ref{fig:lh}.
Figure.~\ref{fig:lh} also shows the numbers of $N_\mathrm{total}$
corresponding to $90\%$ of the accumulated areas below the
likelihood curves, which are then quoted as the upper limits on the
number of signal events at the $90\%$ C.L. The limits are $244$ and
$335$ for the \dswd and \dsstarwd decay modes, respectively.

\section{Systematic Uncertainties}
\label{sec:sysuncer}

Systematic uncertainties in this analysis are divided into two sets.
The dominant one is from the uncertainty of the efficiency corrected 
signal yield. The others are common uncertainties,
including the physics model, electron tracking, electron PID, $E/p$ cut, total
number of \jpsi events, and trigger efficiency, as well as the photon
efficiency and $\mathcal{B}(D_s^{*-} \to D_s^- \gamma)$ for the $D_s^*$
mode.

\subsection{Systematic uncertainty of efficiency corrected signal yield for each channel}
\label{sec:sig:sys:err} The systematic uncertainties caused by
charged and neutral particle reconstruction efficiencies, $K$ and
$\pi$ PID efficiencies, the $\pi^0$ reconstruction efficiency, the
$K_S^0$ reconstruction efficiency, and $D_{s}$ mass resolutions are
all considered together as the systematic error due to the
reconstruction efficiency of the $D_s$. It is the dominant
uncertainty in this analysis and is studied using a control sample
of $\psi(4040) \to D_{s}^{+} D_{s}^{-}$, in which a $478\
\mathrm{pb^{-1}}$ $\psi(4040)$ data sample taken at $4.009 \
\mathrm{GeV}$ is used~\cite{Ablikim:2012ht}. In this study, one
$D_{s}$ is tagged using eight $D_s$ hadronic decays modes, and the
other $D_{s}$ is reconstructed in the same way as in the
\dsordsstarwd ~analysis. The differences of the $D_s$ reconstruction
efficiencies of MC and data are quoted as the systematic
uncertainties in the $D_{s}$ reconstruction and are listed in
Tables.~\ref{tab:dstotaleff} and ~\ref{tab:dsrtotaleff}.  The
uncertainties on the $D_s$ decay branching fractions 
are separated from the reconstruction uncertainty deliberately by
squared subtraction. 


The systematic uncertainty of background shapes is estimated by
varying the shapes of background. These new background shapes are
obtained by smoothing the bin contents of the histograms, that are
extracted from the inclusive MC sample. By convolution with a
Guassian function, we repeat this process till the maximum
difference between the contents of any two adjacent bins is less than
$25\%$.

The systematic uncertainty due to the choice of fitting ranges is determined by
varying the ranges of the $\Umiss$ distributions from
$[-0.2,0.2]$ to $[-0.25,0.25]\ \mathrm{GeV/c^2}$, and the
difference is taken as this systematic uncertainty.

The systematic uncertainty contributions studied above and the uncertainty due to 
MC statistics are summarized in Tables~\ref{tab:dstotaleff} and~\ref{tab:dsrtotaleff}. The total
uncertainty is obtained by summing in quadrature the individual uncertainties
quadratically.

\begin{table}
\footnotesize
\caption{Summary of systematic uncertainties of the efficiency corrected signal yield in the measurement of
  $J/\psi \to D^-_s e^+ \nu_e$ in $\%$.}
\vspace{5mm}
\begin{tabular}{c|c|c|c|c}
\hline \hline
Sources\textbackslash modes  & $K^+ K^- \pi^-$         &   $K^+ K^- \pi^- \pi^0$ & $K^0_S K^-$  & $K^0_S K^- \pi^+ \pi^-$  \\
\hline
Reconstruction $\epsilon$     &       6.8              &      16.2               &       16.6   &        18.6             \\
$\mathcal{B}(D_s^- \to X)$   &    3.9                  &       11.1              &       4.0    &         6.6             \\
Background shape            &        2.3               &    2.4                  &       3.2    &        2.9                \\
Fitting range            &        0.3                  &    0.4                  &       0.5    &        0.6                \\
MC statistic            &        0.7                   &    1.2                  &       0.6    &        0.9                \\
\hline
Total                    &        8.2                  &      19.8               &       17.4   &        20.0           \\
\hline \hline
\end{tabular}
\label{tab:dstotaleff}
\end{table}

\begin{table}
\footnotesize
\caption{Summary of systematic uncertainties of the efficiency corrected signal yield in the measurement of
  \dsstarwd in $\%$.}
\vspace{5mm}
\begin{tabular}{c|c|c|c|c}
\hline \hline
Sources\textbackslash modes  & $K^+ K^- \pi^-$  & $K^+ K^- \pi^- \pi^0$& $K^0_S K^-$      & $K^0_S K^- \pi^+ \pi^-$  \\
\hline
Reconstruction $\epsilon$     &       6.8       &      16.2            &       16.6       &        18.6             \\
$\mathcal{B}(D_s^-\to X)$     &       3.9       &       11.1           &       4.0        &         6.6             \\
Background shape            &     2.5           &    2.5               &       2.7        &        3.2              \\
Fitting range            &        0.2           &    0.6               &       0.4        &        0.4              \\
MC statistic            &         1.0           &     1.9              &      0.9         &         1.4             \\
\hline
Total                    &        8.3           &      20.0            &        17.4      &           20.1           \\
\hline \hline
\end{tabular}
\label{tab:dsrtotaleff}
\end{table}

\subsection{Common uncertainties}
The difference of the efficiencies based on phase space and
the new generator used in this analysis is taken as the systematic
uncertainty of the physics model.

The systematic uncertainty of the resolutions has been estimated by
smearing the MC simulations. The simulation of the photon
reconstruction has been studied with a control sample of the 
well-understood decays $J/\psi \to \rho^0 \pi^0$ in
Ref.~\cite{Ablikim:2010zn}, and we smear the resolution of the
photon energy deposited in the EMC at the $1\%$ level by
a convolution with a Gaussian function.
For the tracks from charged particles, we smear the helix parameters of
each track as described in Ref.~\cite{Ablikim:2012pg}. 
The difference in the final yields between before and after smearing is 
taken as the systematic uncertainty.  The variable $\Umiss$ is associated 
with the energy and momentum resolutions of detected tracks. Thus, 
the systematic uncertainty of the signal shape has been taken into
account implicitly. 

The electrons from the signal are in a low momentum region, which
causes a systematic uncertainty of $2.1\%$ in the MDC tracking
efficiency and $1.0\%$ in the PID efficiency~\cite{Ablikim:2013wfg}.
A radiative Bhabha sample, normalized with respect to the momentum,
is used as a control sample to estimate the systematic uncertainty
caused by the $E/p$ requirement, i.e. $0.80<E/p<1.05$. The
difference in efficiency between the MC simulation and the data is quoted as the
systematic uncertainty caused by this requirement.
Since the electron momentum in the \dsstarwd decay is lower,
the uncertainty caused by the $E/p$ requirement of \dsstarwd is
larger than that of \dswd correspondingly.

The total number of $J/\psi$ events is determined by using $J/\psi$ inclusive
decays~\cite{Ablikim:2012cn}, and the value $1.2\%$ is quoted as the
systematic uncertainty of the total number of $J/\psi$ events.

According to Ref.~\cite{Berger:2010my}, the trigger efficiency is
very high since there are four to six tracks from charged particles in addition to
possible neutral particles within the barrel regions in the final
states. Therefore, the systematic uncertainty of the trigger efficiency is
negligible.

Since the $D_s^*$ mesons are only reconstructed by ${D^{*}_{s}}^{-}
\to D_{s}^{-} \gamma$, we deal with most of the systematic
uncertainties of \dsstarwd in the same way as those of \dswd, and
with two additional uncertainties in $D^*_s$ than in the $D_s$ mode.
One is a $1\%$ uncertainty from the additional photon detecting
efficiency~\cite{Ablikim:2011kv}. The other one is the input
branching fraction $\mathcal{B}({D^{*}_{s}}^{-} \to D_{s}^{-}
\gamma)$ in MC simulation. Since the world average value is $(94.2
\pm 0.7)\%$~\cite{Beringer:1900zz}, this leads to a $0.7\%$
uncertainty.
All of the common systematic uncertainties are listed in
Table~\ref{tab:totalerr}.

\begin{table}
\caption{Summary of common systematic uncertainties in the
measurement of \dswd and \dsstarwd.} \vspace{5mm}
\begin{tabular}{c|c|c}
\hline \hline
Source                                 &  \dswd (\%)                   &      \dsstarwd  (\%) \\
\hline
Physics model                          &         0.9                   &        0.8              \\
Resolutions                            &         1.6                   &         1.8             \\
$e$ tracking                           &         2.1                   &         2.1            \\
$e$ PID                                &         1.0                   &         1.0             \\
$E/p$ cut                              &         0.6                   &          1.7            \\
Photon efficiency                      &           -                   &           1.0             \\
$\mathcal{B}(D_s^{*-} \to D_s^- \gamma)$ &         -                   &           0.7              \\
$J/\psi$ events                        &         1.2                   &           1.2               \\
Trigger                                &    Negligible                 &     Negligible         \\
\hline
Total                                  &       3.3                     &           3.9           \\
\hline \hline
\end{tabular}
\label{tab:totalerr}
\end{table}

\subsection{Upper limit calculation}

Taking the systematic uncertainties into account, the upper limits on
the branching fractions are calculated using
\begin{equation}
\mathcal{B} < \frac{{N^\mathrm{up~\prime}_\mathrm{total}}}{(1-\sigma_\mathrm{common}^\mathrm{sys})N_{J/\psi}}
\ \ , \label{equ:aft:br}
\end{equation}
where $N^\mathrm{up~\prime}_\mathrm{total}$ is the corrected $N_\mathrm{total}^\mathrm{up}$ after considering
the systematic uncertainties of the signal efficiency, as described below, and
$\sigma_\mathrm{common}^\mathrm {sys}$ is the total common systematic uncertainty.

From Eqs.~(\ref{eq:likelihood}) and~(\ref{eq:90likelihood}), $N_\mathrm{total}^\mathrm{up}$ depends on 
the signal efficiencies of all decay channels in a complex way, and there is no 
simple analytic method to calculate the final effect due to those efficiency 
uncertainties. To study this dependence, we obtain an $N_\mathrm{total}^\mathrm{up}$ 
distribution by sampling each signal efficiency by a Gaussian function of which mean 
value and standard deviation are set as the normal signal efficiency and the systematic 
uncertainty obtained before, respectively. This new $N_\mathrm{total}^\mathrm{up}$ 
distribution can be described by a Gaussian function. A sum of the mean 
value ($\bar{N}_\mathrm{total}^\mathrm{up}$) and one standard deviation ($\sigma_\mathrm{total}$) 
of this Gaussian function is quoted as the ${N^\mathrm{up~\prime}_\mathrm{total}}$. 
All the numerical results are summarized in Table~\ref{tab:bfc}.


\begin{table}
\caption{Upper limits of the branching fractions of \dswd  and 
  \dsstarwd after considering the systematic uncertainties.}
\vspace{5mm}
\begin{tabular}{ccc}
\hline \hline
                         &     \dswd       &     \dsstarwd   \\
\hline
$\bar{N}_\mathrm{total}^\mathrm{up}$     &      244        &        335      \\
$\sigma_\mathrm{total}$         &       31        &         43      \\
${N^\mathrm{up~\prime}_\mathrm{total}}$  &      275        &        378      \\
$\sigma^\mathrm{sys}_\mathrm{common}$  &      $3.3\%$    &        $3.9\%$  \\
$N_{J/\psi}$             &     \multicolumn{2}{c}{$2.25\times 10^8$}  \\
\hline
$\mathcal{B}(90\% \mathrm{C.L.})$ & $<1.3\times10^{-6}$ &  $<1.8\times10^{-6}$   \\
\hline \hline
\end{tabular}
\label{tab:bfc}
\end{table}

\section{Summary}
With a sample of $2.25 \times 10^{8}$ \jpsi events collected with
the BESIII detector, we have searched for the weak decays \dswd and
\dsstarwd. No significant excess of signal is observed. At the 90\%
C.L., the upper limits of the branching fractions are:
$\mathcal{B}(J/\psi \to D^{-}_{s}e^{+}\nu_{e}+c.c.)< 1.3 \times
10^{-6}$ and $\mathcal{B}(J/\psi \to
{D^{*}_{s}}^{-}e^{+}\nu_{e}+c.c.) < 1.8\times10^{-6}$. The upper
limit on the branching fraction $\mathcal{B}(J/\psi \to
D^{*-}_{s}e^{+}\nu_{e}+c.c.)$ is set for the first time and the
upper limit on the branching fraction $\mathcal{B}(J/\psi \to
D^{-}_{s}e^{+}\nu_{e}+c.c.)$ is 30 times more strict than the
previously result~\cite{Beringer:1900zz}.
The results are within the SM prediction, but more data will
be helpful to test the branching fraction of semileptonic decays
of the \jpsi to the order of $10^{-8}$. The results would
also be applied to constrain the parameter spaces of some
BSM models if direct calculations of these processes are carried out
in the future. 

\vspace{10mm}

\section*{Acknowledgment}
The BESIII collaboration thanks the staff of BEPCII and the IHEP
computing center for their strong support. This work is supported in
part by National Key Basic Research Program of China under Contract
No.~2015CB856700; Joint Funds of the National Natural Science
Foundation of China under Contracts Nos.~11079008, 11179007, 11179014, U1332201;
National Natural Science Foundation of China (NSFC) under Contracts
Nos.~10625524, 10821063, 10835001, 11125525, 11235011, 11335008; the
Chinese Academy of Sciences (CAS) Large-Scale Scientific Facility
Program; CAS under Contracts Nos.~KJCX2-YW-N29, KJCX2-YW-N45; 100
Talents Program of CAS; German Research Foundation DFG under Contract
No.~Collaborative Research Center CRC-1044; Istituto Nazionale di
Fisica Nucleare, Italy; Ministry of Development of Turkey under
Contract No.~DPT2006K-120470; National Natural Science Foundation of
China (NSFC) under Contract No.~11275189; Russian Foundation for Basic
Research under Contract No.~14-07-91152; U. S. Department of Energy
under Contracts Nos.~DE-FG02-04ER41291, DE-FG02-05ER41374,
DE-FG02-94ER40823, DESC0010118; U.S. National Science Foundation;
University of Groningen (RuG) and the Helmholtzzentrum fuer
Schwerionenforschung GmbH (GSI), Darmstadt; WCU Program of National
Research Foundation of Korea under Contract No.~R32-2008-000-10155-0.


\begin{thebibliography}{99}


\bibitem{Sanchis:1992pv}
  M.~A.~Sanchis,
  Phys.\ Lett.\ B {\bf 280}, 299 (1992).

\bibitem{SanchisLozano:1993ki}
  M.~A.~Sanchis-Lozano,
  Z.\ Phys.\ C {\bf 62}, 271 (1994).

\bibitem{Wang:2007ys}
  Y.~-M.~Wang, H.~Zou, Z.~-T.~Wei, X.~-Q.~Li and C.~-D.~Lu,
  Eur.\ Phys.\ J.\ C {\bf 54}, 107 (2008).

\bibitem{xm.zhang2001}
  X.-M, Zhang, High Energy Physics and Nuclear Physics {\bf 25}, 461 (2001).


\bibitem{Li:2012vk}
  H.~-B.~Li and S.~-H.~Zhu,
  Chin.\ Phys.\ C {\bf 36}, 932 (2012).



\bibitem{Datta:1998yq}
  A.~Datta, P.~J.~O'Donnell, S.~Pakvasa and X.~Zhang,
  Phys.\ Rev.\ D {\bf 60}, 014011 (1999).

\bibitem{hill:1995} C. Hill, Phys. Lett. B {\bf 345}, 483 (1995).

\bibitem{Ablikim:2006qt}
  M.~Ablikim {\it et al.}  [BES Collaboration],
  Phys.\ Lett.\ B {\bf 639}, 418 (2006).


\bibitem{Ablikim:2012cn}
  M.~Ablikim {\it et al.}  [BESIII Collaboration],
  Chin.\ Phys.\ C {\bf 36}, 915 (2012).


\bibitem{Ablikim:2012ht}
  M.~Ablikim {\it et al.}  [BESIII Collaboration],
  Phys.\ Rev.\ D {\bf 86}, 071101 (2012).


\bibitem{Ablikim:2009aa}
  M.~Ablikim {\it et al.}  [BESIII Collaboration],
  Nucl.\ Instrum.\ Meth.\ A {\bf 614}, 345 (2010).

\bibitem{Agostinelli:2002hh} 
  S.~Agostinelli {\it et al.}  [GEANT4 Collaboration],
  Nucl.\ Instrum.\ Meth.\ A {\bf 506}, 250 (2003).

\bibitem{Abakumova:2011rp}
  E.~V.~Abakumova, M.~N.~Achasov, V.~E.~Blinov{\it et al.},
  Nucl.\ Instrum.\ Meth.\ A {\bf 659}, 21 (2011).


\bibitem{kkmc}
  S.~Jadach, B.~F.~L.~Ward and Z.~Was, Comp. Phys. Commu. {\bf 130}, 260 (2000);
  S.~Jadach, B.~F.~L.~Ward and Z.~Was, Phys. Rev. D {\bf 63}, 113009 (2001).

\bibitem{evtgen}
  D.~J.~Lange, Nucl. Instrum. Meth. A {\bf 462},152 (2001).


\bibitem{Beringer:1900zz}
  K.A. Olive {\it et al.} [Particle Data Group], Chin. Phys. C, {\bf 38}, 090001 (2014).

\bibitem{lundcharm}
  R.~G.~Ping,
  Chin. Phys. C {\bf 32}, 599 (2008).


\bibitem{Gaiser:1982yw}
  J.~Gaiser,
  SLAC Stanford - SLAC-255 (82,REC.JUN.83) 194p.

\bibitem{Cranmer:2000du}
  K.~S.~Cranmer,
  Comput.\ Phys.\ Commun.\  {\bf 136}, 198 (2001).


\bibitem{Ablikim:2010zn}
  M.~Ablikim {\it et al.}  [BESIII Collaboration],
  Phys.\ Rev.\ D {\bf 81}, 052005 (2010).

\bibitem{Ablikim:2012pg}
  M.~Ablikim {\it et al.}  [BESIII Collaboration],
  Phys.\ Rev.\ D {\bf 87}, 012002 (2013).

\bibitem{Ablikim:2013wfg}
  M.~Ablikim {\it et al.}  [BESIII Collaboration],
  Phys.\ Rev.\ D {\bf 87}, 092011 (2013).


\bibitem{Berger:2010my}
  N.~Berger, K.~Zhu, Z.-A.~Liu, D.-P.~Jin, H.~Xu, W.-X.~Gong, K.~Wang, G.-F.~Cao, 
  Chin.\ Phys.\ C {\bf 34}, 1779 (2010).

\bibitem{Ablikim:2011kv}
  M.~Ablikim {\it et al.}  [BESIII Collaboration],
  Phys.\ Rev.\ D {\bf 83}, 112005 (2011).



\end{thebibliography}
\end{document}